\documentclass[aps,pra,amsmath,amssymb,superscriptaddress,twocolumn,10pt,longbibliography]{revtex4-1}

\usepackage{times,txfonts}
\usepackage{nicefrac}
\usepackage{xr-hyper}
\usepackage[colorlinks=true,linkcolor=blue,urlcolor=blue,citecolor=magenta,pdfusetitle]{hyperref}
\usepackage{epsfig} 
\usepackage{subfigure}
\usepackage{physics}
\usepackage{bm}
\usepackage{bbm}
\usepackage{xcolor}
\usepackage{mathrsfs}%
\usepackage{subfigure}

\def\bra#1{\langle{#1}|}
\def\ket#1{|{#1}\rangle}

\renewcommand{\vec}{\bm}

\def\e{{\mathrm{e}}}

\def\ii{{\mathrm{i}}}

\begin{document}

\pagecolor{white}

\title{A digital Rydberg simulation of dynamical quantum phase transitions in the Schwinger model}

\author{Domenico Pomarico}
\affiliation{Dipartimento di Fisica, Universit\`{a} di Bari, I-70126 Bari, Italy}
\affiliation{INFN, Sezione di Bari, I-70125 Bari, Italy}

\author{Federico Dell'Anna}
\affiliation{Dipartimento di Fisica e Astronomia, Universit\`{a}  di Bologna, I-40127 Bologna, Italy}
\affiliation{INFN, Sezione di Bologna, I-40127 Bologna, Italy}

\author{Riccardo Cioli}
\affiliation{Dipartimento di Fisica e Astronomia, Universit\`{a}  di Bologna, I-40127 Bologna, Italy}
\affiliation{INFN, Sezione di Bologna, I-40127 Bologna, Italy}

\author{Saverio Pascazio}
\affiliation{Dipartimento di Fisica, Universit\`{a} di Bari, I-70126 Bari, Italy}
\affiliation{INFN, Sezione di Bari, I-70125 Bari, Italy}

\author{Francesco V. Pepe}
\affiliation{Dipartimento di Fisica, Universit\`{a} di Bari, I-70126 Bari, Italy}
\affiliation{INFN, Sezione di Bari, I-70125 Bari, Italy}

\author{Paolo Facchi}
\affiliation{Dipartimento di Fisica, Universit\`{a} di Bari, I-70126 Bari, Italy}
\affiliation{INFN, Sezione di Bari, I-70125 Bari, Italy}

\author{Elisa Ercolessi}
\affiliation{Dipartimento di Fisica e Astronomia, Universit\`{a}  di Bologna, I-40127 Bologna, Italy}
\affiliation{INFN, Sezione di Bologna, I-40127 Bologna, Italy}

\begin{abstract}
We present the simulation of the quench dynamics of the $\mathbb{Z}_3$ Schwinger model, that describes an approximation of one-dimensional Quantum Electrodynamics, on a digital noisy Rydberg atom platform, aiming at the observation of multiple dynamical quantum phase transitions. In order to reach long-time dynamics, we exploit an enconding dictated by the symmetries,  combined with a circuit compression procedure. 
We focus on a quench that evolves the Dirac vacuum by means of a Hamiltonian depending on a negative mass parameter. This leads to resonant Rabi oscillations between the Dirac vacuum and  mesonic states. The population concentration exhibits oscillations with negligible fluctuations of detuned states also with the inclusion of combined noise sources, from which we can clearly detect multiple dynamical phase transitions.
\end{abstract}


\maketitle


\section{Introduction}\label{intro}
We address the problem of simulations of lattice gauge models, more specifically of Quantum Electrodynamics (QED), on noisy intermediate-scale quantum (NISQ) devices. 
As for generic gauge theories, lattice QED requires the implementation of the constraint projecting onto the physical subspace which consists of gauge invariant states. This relates such model  with the category of kinetically constrained models~\cite{papic2018, pxp2020, ergo2023}, which are known to be endowed with interesting many-body phenomena like quantum scarring, that are preserved through linear gauge protection~\cite{zeno, zeno2, zeno3, halimeh2022stabilizing}, and Dynamical Quantum Phase Transitions 
(DQPTs) corresponding to the appearance of non-analyticities  in the free-energy density after a quench ~\cite{mueller2023quantum, dqpt, dqpt2, dqpt_QLM, dqpt_QLM2, osborne2023probing, corps2024unifying}, which would be very interesting to probe via experimental realizations~\cite{Wang_2023,dqpt}.

A proper implementation of quantum computing on NISQ devices can be targeted by means of multiple strategies. IBM Quantum platform exploits compiler capabilities to optimize circuits, like depth compression via template matching~\cite{qiskit_opt, Qiskit}. Nonetheless such procedures ruled at a software level are accompanied by those referred to the hardware level. In the aforementioned superconducting framework, the state of transmon qubits~\cite{cirq, amazon} is elaborated by means of maser pulses, tuned with respect to targeted two-level transition. A crucial optimization of the experimental protocol resides in the achievement of short pulse duration in order to apply a certain logical gate. An example of these minimization tricks is represented by virtual $Z$ gates, consisting in an instantaneous phase shift able to implement a $Z$ rotation up to an overall phase shift~\cite{virtualZ}. The same hardware optimization recipe is shared by Rydberg atoms setups, where laser pulses control atomic optical transitions. The probability to undergo interaction with the environment leading to a decoherent evolution is clearly enhanced for increasing depth of the circuit. Superconducting devices are endowed with much shorter duration of physical implementation for each gate, balanced by a higher error rate, especially affecting double qubits cases, thus motivating procedures for circuit compression~\cite{qiskit_opt}. On the other hand neutral Rydberg atoms quantum computing guarantees higher gate fidelity~\cite{rydberg_fid1, rydberg_fid2, doultsinos2024quantumgatesdistantatoms}, but requiring a longer time in the experimental execution. 

In this paper, we choose to work on a Rydberg atom platform. We make use of our digital simulation proposal for the observation of multiple DQPTs associated with resonant Rabi oscillations~\cite{dqpt_QLM, dqpt_QLM2}, characterizing the quenched dynamics of two sites described by a $\mathbb{Z}_3$ QED model, in the situation in which the Dirac vacuum is suddenly evolved by means of a Hamiltonian with a negative mass parameter, a situation in which mesonic ground states are expected to play a crucial role~\cite{Ising}. It is worth noticing that dynamics is not trivial even for two lattice sites, where an analytical solution is lacking and numerical diagonalization is needed.

We first design the digital evolution generated by the QED Hamiltonian in $1+1$ dimensions to probe DQPTs, by exploiting the fact that we can introduce a specific quantum-classical strategy for the encoding~\cite{ibm_qed,entropy2} within each conservation law sector. This is useful to avoid any migration driven by errors towards undesired degrees of freedom and represents a necessary step to study the evolution of the system in proximity of a DQPT, since the latter is very strongly affected by noise~\cite{QLM,QLMopen,dqpt_ising,entropy}.  Simulations of these devices are studied through the Python package \texttt{pulser}~\cite{pulser}, which allows to take into account most typical noise sources.  Also, an essential improvement, required to include long-time dynamics and consequently the realization of multiple transitions, is ensured by the exploitation of the Python package \texttt{qiskit} for the circuit depth compression~\cite{qiskit_opt,3cnot}.

The structure of the paper is the following. In Section~\ref{model}, we review QED on a one-dimensional lattice, discretizing the continuous gauge group $U(1)$ with the cyclic group $\mathbb{Z}_n$ and introducing staggered fermions to recover the Schwinger model~\cite{PhysRev.82.664, PhysRev.128.2425} . We then discuss how to exploit parity symmetry to encode physical degrees of freedom and describe the quench protocol. Rydberg platform is presented in Section~\ref{simulation}, where a brief tutorial about available pulses is followed by the description of Trotterized time evolution. We discuss pulse sequence optimization and benchmark the resilience of our protocol with respect to noise sources. In Section~\ref{results}, we perform the (noisy) simulation of the full quench protocol for a two-site $\mathbb{Z}_3$  Schwinger model in different parameter regimes,  where we can achieve long time evolutions capable of capturing multiple DQPTs. Finally, we draw our conclusions.

\section{The Schwinger model}\label{model}

QED in a one-dimensional (1D) lattice, known equivalently as the Schwinger model~\cite{PhysRev.82.664, PhysRev.128.2425}, was the first gauge theory whose continuous-time dynamics was simulated on a real quantum computing setup~\cite{ionq, ionq2}.  In this model,  spinless fermions, endowed with mass $\pm m$ and charge $g$, interact with the electromagnetic field described by the abelian $U(1)$ gauge theory, whose unique degree of freedom is a scalar electric component, with any magnetic contribution missing due to the absence of plaquettes in the lattice. An anticommuting spinor field $\psi_x$ is associated with each lattice site $x$, joined by links of length $a$, which host the electric field $E_{x,x+1}$ and the vector potential $A_{x,x+1}$. The gauge connection $U_{x,x+1} = \e^{\ii a A_{x,x+1}}$ allows us to introduce the finite-differences covariant derivative which yields the hopping term in the Hamiltonian for the considered periodic lattice~\cite{QLM,QLMopen,entropy}
\begin{eqnarray}
\mathcal{H}(m, J, g) &=& - \frac{\ii J}{2} \sum_{x=0}^{N-1} \left( \psi_x^\dagger U_{x,x+1} \psi_{x+1} - \mbox{H.c.} \right) \nonumber\\
& &+ m \sum_{x=0}^{N-1} (-1)^x \psi_x^\dagger \psi_x + \frac{g^2}{2 J} \sum_{x=0}^{N-1} E_{x,x+1}^2,
\label{Eq::HmJ} 
\end{eqnarray}
where the identification $N\equiv 0$ for site labels is used and $J=1/a$. The model exploits staggered (Kogut-Susskind) fermions to solve the doubling problem~\cite{stagg}. Then, the Dirac sea consists of odd-$x$ sites hosting a negative mass component, while even sites are empty. This Dirac vacuum state coincides with the system ground state in the limit $m \rightarrow \infty$.

The simulation of gauge degrees of freedom requires a discretization of the $U(1)$ parameter space, implemented via a $\mathbb{Z}_n$ model, which is based on the use of permutation matrices in place of gauge connections~\cite{Zn}. This discrete approximation, which differs from the quantum link model, where spin ladder operators are obtained via algebra truncation~\cite{QLM,QLMopen,dqpt_QLM,dqpt_QLM2}, is known to have the same quantum phase diagram of the $U(1)$ Schwinger model~\cite{Ising}.  Here, the gauge degrees of freedom are described by an $n$-dimensional Hilbert space $\mathscr{H}_n$ with orthonormal basis $\{|e_k\rangle\}_{k=0}^{n-1}$ such that
\begin{eqnarray}
  U_{x,x+1} \ket{e_k} = \ket{e_{k+1}}, \quad U_{x,x+1} \ket{e_{n-1}} = \ket{e_0}, \quad E_{x,x+1} \ket{e_k} = e_k \ket{e_k},\nonumber\\
\end{eqnarray}
where the $\mathbb{Z}_n$ cyclic group elements are represented by 
 \begin{equation}
  e_k = \sqrt{\frac{2\pi}{n}} \left(k-\frac{n-1}{2}\right).
\end{equation}

It is now possible to map the spinor field into spin variables through a Jordan-Wigner transformation~\cite{ionq,entropy}, so that the Hamiltonian with rescaled parameters $\xi = J^2 / g^2$, $\mu = m J / g^2$ reads
\begin{eqnarray} 
\tilde{\mathcal{H}}(\xi, \mu) &=& \xi \sum_{x=0}^{N-1} \left( \sigma_x^- U_{x,x+1} \sigma_{x+1}^+ + \mbox{H.c.} \right) 
\nonumber\\
& &- \mu \sum_{x=0}^{N-1} (-1)^x Z_x + \sum_{x=0}^{N-1} E_{x,x+1}^2,
\label{Eq::Hximu}
\end{eqnarray}
with $\sigma_x^{\pm} = \frac{1}{2}(X_x \pm \ii Y_x)$, where $\bm{\sigma}_x=(X_x,Y_x,Z_x)$ is the vector of spin-$1/2$ Pauli operators associated to site $x$. In the total Hilbert space $\mathscr{H}=  \bigotimes_{x=0}^N( \mathbb C^2 \otimes \mathscr{H}_n)$ physical states are defined with respect to the Gauss-law constraint $G_x \ket{\phi} = 0$ at all sites $x$, with
\begin{equation}
G_x = \sqrt{\frac{n}{2\pi}} \left(E_{x,x+1} - E_{x-1,x}\right) - \psi_x^\dagger \psi_x - \frac{(-1)^x - 1}{2},
\end{equation}
determining the states $\ket{\phi}$ that belong to the gauge-invariant subspace $\mathscr{H}_G$.

\begin{figure}
	\subfigure[]{\includegraphics[width=0.32\linewidth]{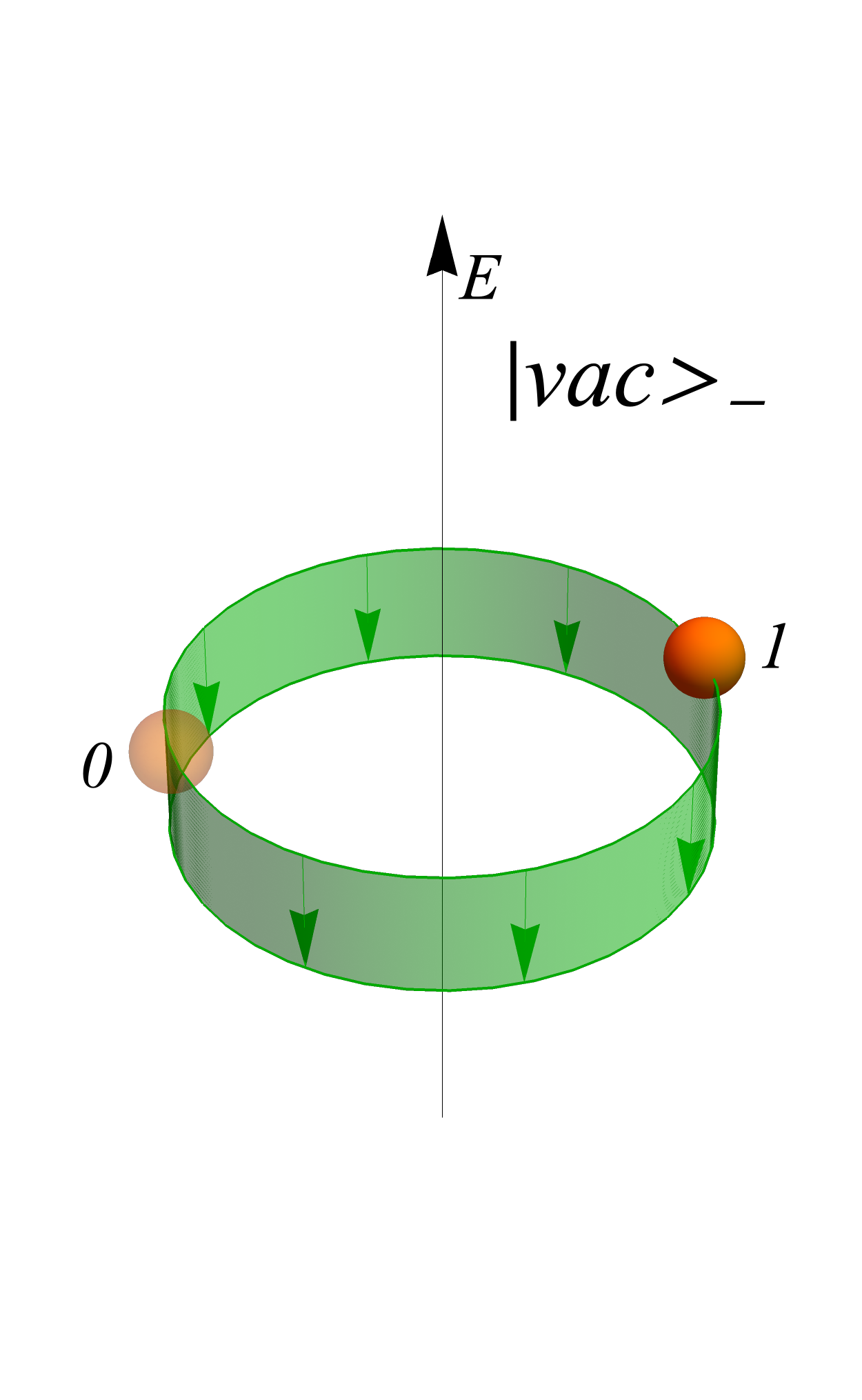}}
	\subfigure[]{\includegraphics[width=0.32\linewidth]{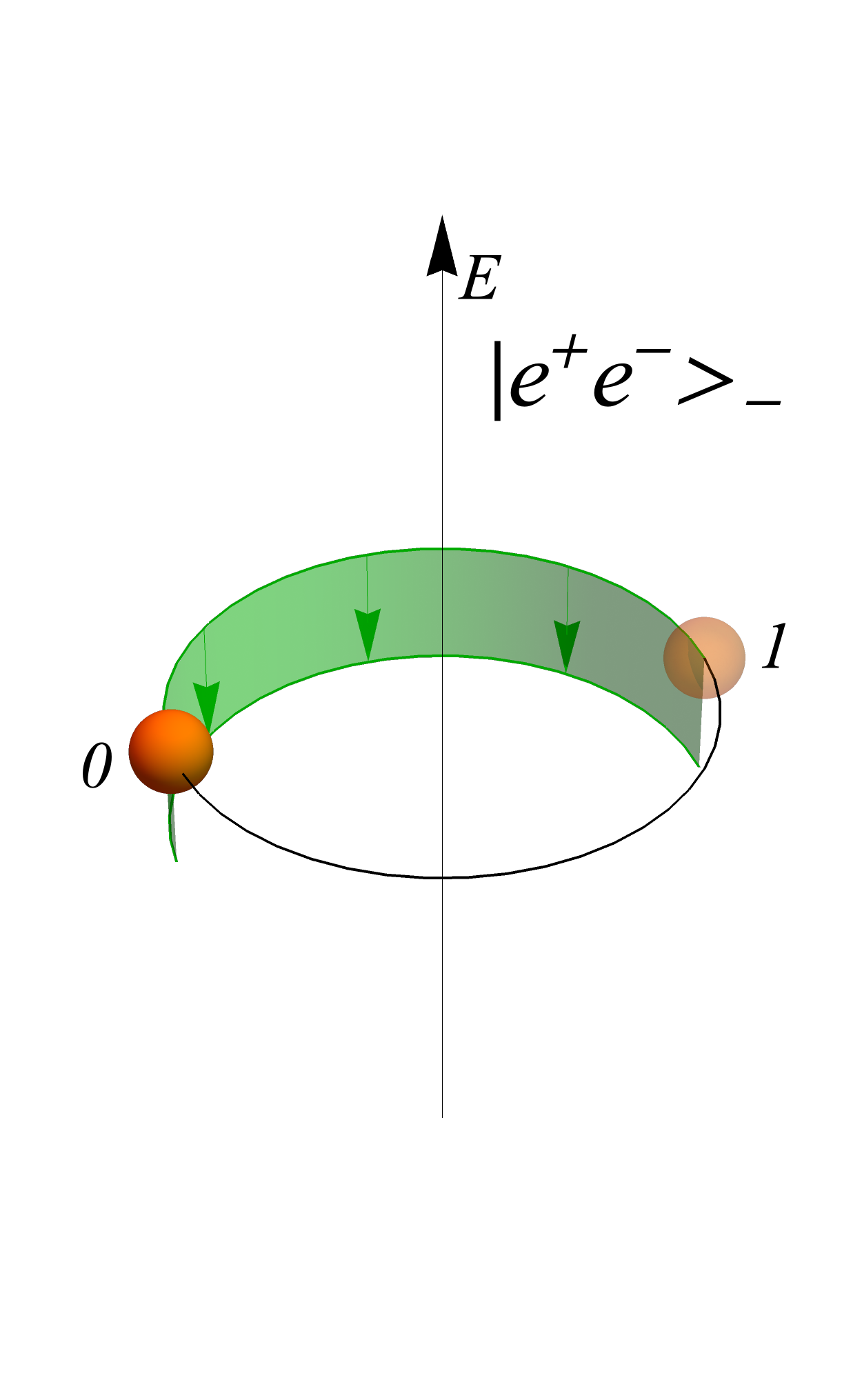}}
	\subfigure[]{\includegraphics[width=0.32\linewidth]{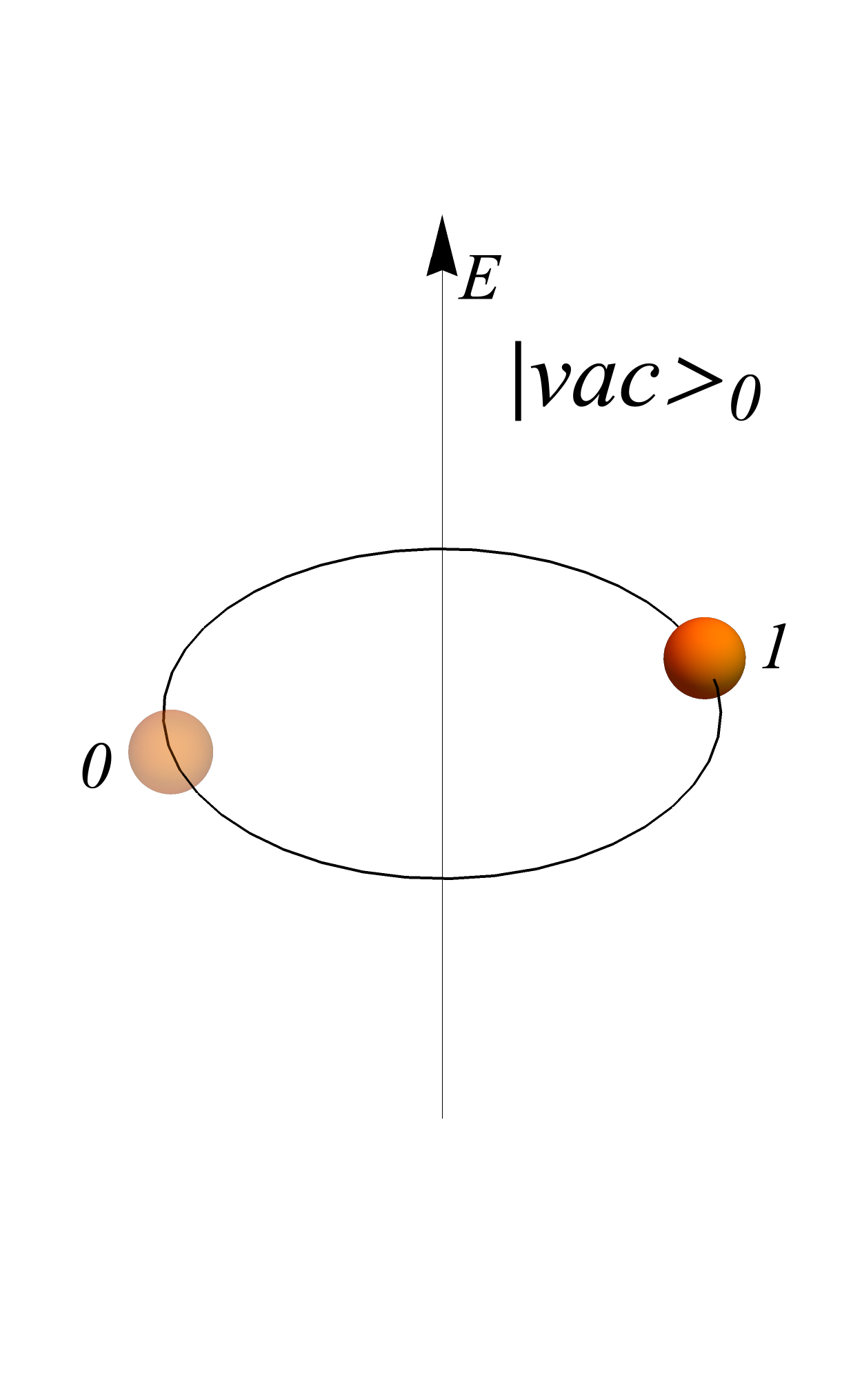}}
	\subfigure[]{\includegraphics[width=0.32\linewidth]{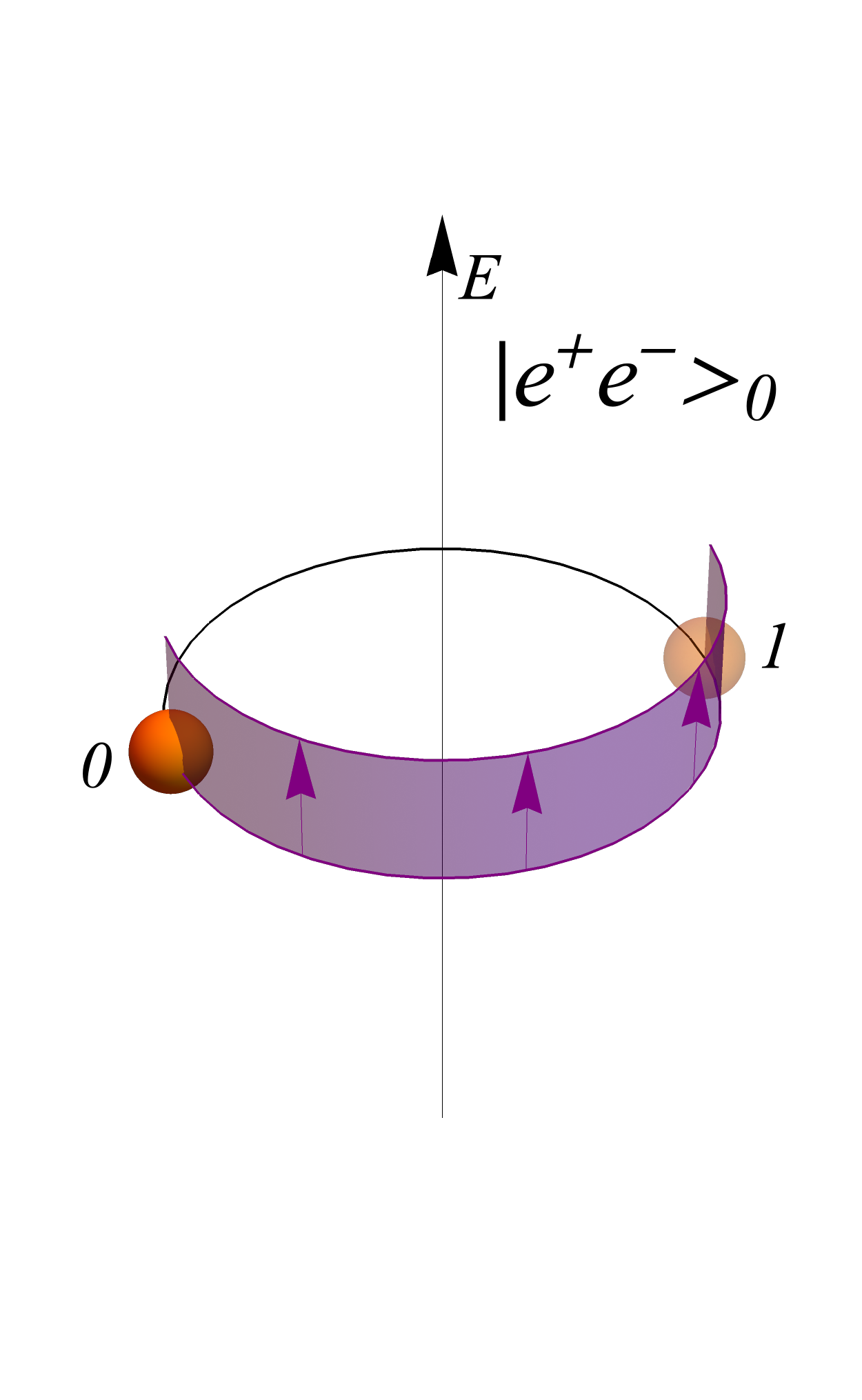}}
	\subfigure[]{\includegraphics[width=0.32\linewidth]{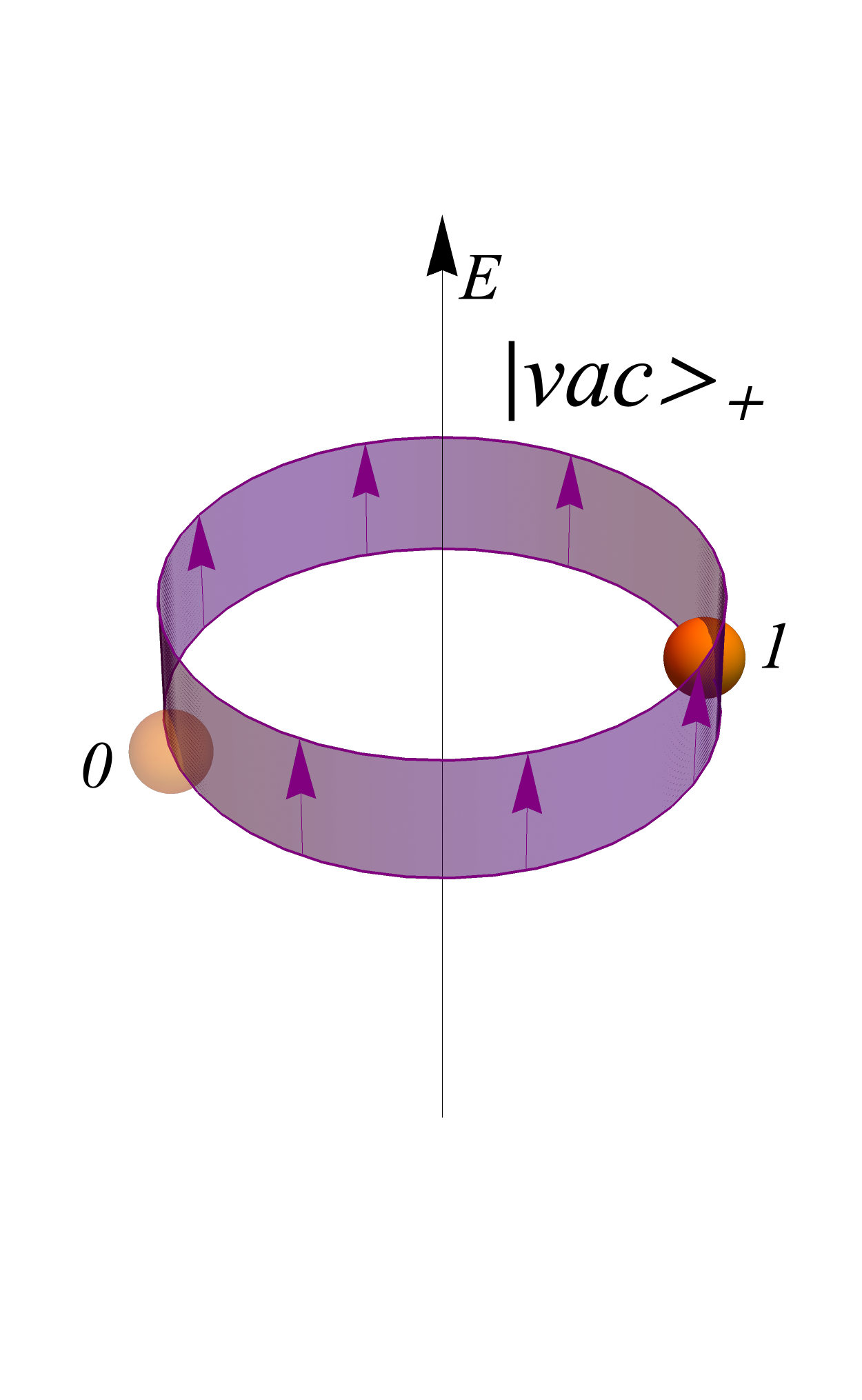}}
	\subfigure[]{\includegraphics[width=0.32\linewidth]{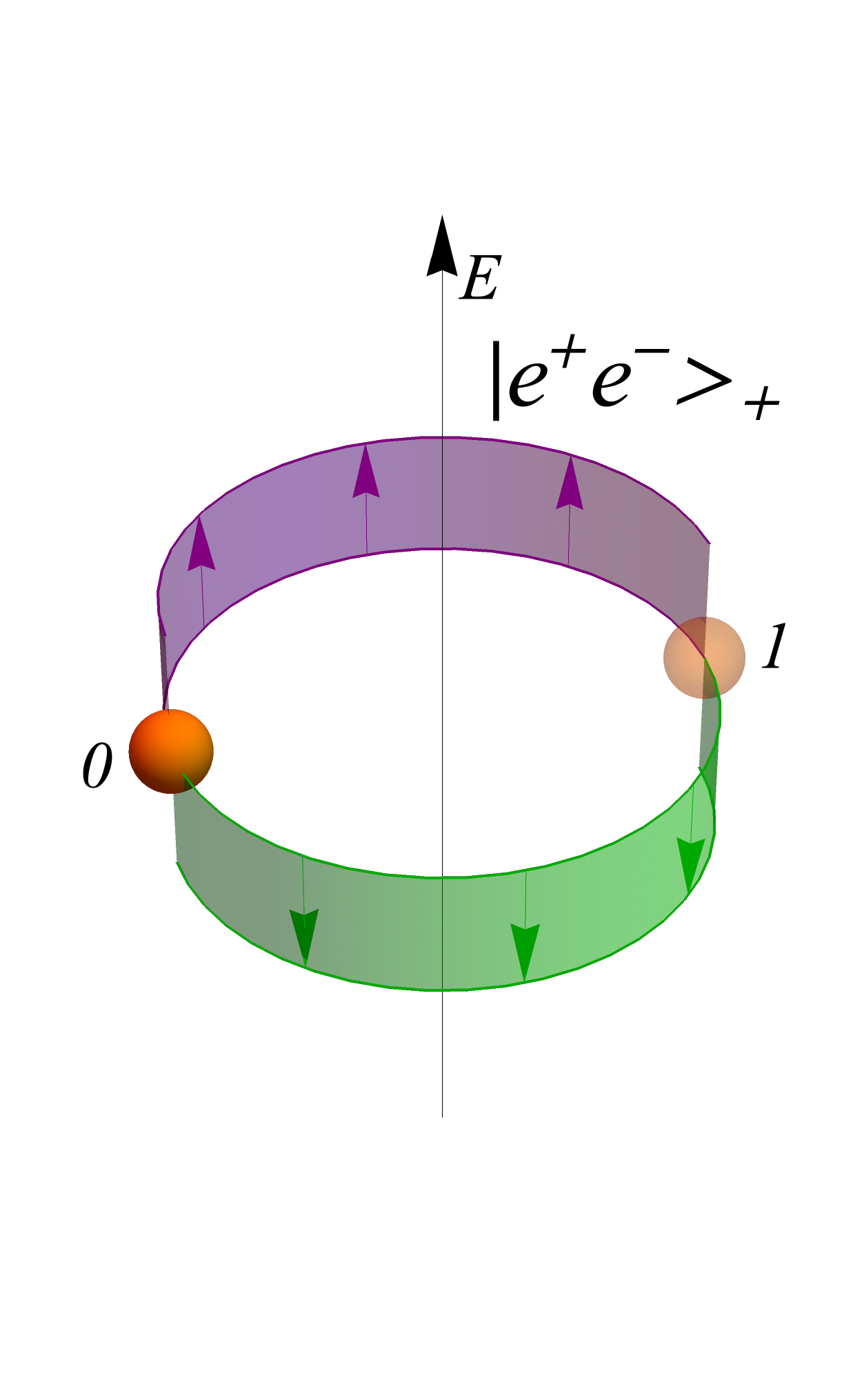}}
    \caption{Representation of physical states for 2 matter sites in the $\mathbb{Z}_3$ Schwinger model, listed according to the notation $\{ \ket{\mathrm{vac}}_{\varepsilon_0}, \ket{e^+ e^-}_{\varepsilon_0} \}$ corresponding to an electric field $\varepsilon_0 = -1, 0, +1$ on the  link from $x=1$ to $x=0$.} \label{fig:phys_2sites}
\end{figure}

Here, we consider a periodic lattice of $N = 2$ sites with an equal number of links and a $\mathbb{Z}_3$ gauge group, yielding $e_k = \sqrt{\frac{2\pi}{3}} (k-1)$ with $k=0,1,2$, whose physical states are represented in Fig.~\ref{fig:phys_2sites}. There are three Dirac vacua $|\mathrm{vac}\rangle_{\varepsilon_0}$, shown in panels (a)-(c)-(e),  where only the site $x=1$ is occupied while the site $x=0$ is empty and corresponding to the three possible values of the electric fields $\varepsilon_0= -1,0,+1$ on the  link from $x=1$ to $x=0$. There are also three mesonic states $|e^+e^-\rangle_{\varepsilon_0}$, shown in panels (b)-(d)-(f), which are obtained from the the associated vacuum $\ket{\mathrm{vac}}_{\varepsilon_0}$ by hopping the fermion from site $x=1$ to $x=0$. In general, the physical subspace dimension for a periodic lattice scales linearly with the number of electric field states per link, namely $\mathrm{dim} \ \mathscr{H}_G = n \binom{N}{N/2}$, where $\binom{N}{{N}/{2}}$ is the dimension for an open boundary case with a set background electric field $\varepsilon_0$~\cite{Ising}.

In the following, we will exploit a quantum-classical approach in order to achieve an encoding optimization by means of the parity operator~\cite{ibm_qed,entropy2}
\begin{eqnarray}
P &=& \left( \ket{e^+ e^-}_{0\ -}\bra{e^+ e^-} + \ket{\mathrm{vac}}_{- \ +}\bra{\mathrm{vac}} + \mbox{H.c.} \right) 
\nonumber\\
& & + \;\ket{e^+ e^-}_{+\ +}\bra{e^+ e^-} + \ket{\mathrm{vac}}_{0 \ 0}\bra{\mathrm{vac}},
\end{eqnarray}
restricted to $\mathscr{H}_G$, verifying the symmetry property $P \tilde{\mathcal{H}} P = \tilde{\mathcal{H}}$ of the Hamiltonian~\eqref{Eq::Hximu}. We can use the basis of  the parity operator eigenvectors,
\begin{align}
    \ket{\psi_1^+} = & \ \ket{\mathrm{vac}}_0, \qquad\quad\quad\quad\;\;\ \ket{\psi_2^+} = \frac{\ket{e^+ e^-}_0 + \ket{e^+ e^-}_-}{\sqrt{2}}, \label{Eq::phys1}\\ 
    \ket{\psi_3^+} = & \ \frac{\ket{\mathrm{vac}}_+ + \ket{\mathrm{vac}}_-}{\sqrt{2}}, \qquad \ket{\psi_4^+} = \ket{e^+ e^-}_+, \label{Eq::phys2}\\ 
    \ket{\psi_1^-} = & \ \frac{\ket{e^+ e^-}_0 - \ket{e^+ e^-}_-}{\sqrt{2}}, \quad \ket{\psi_2^-} = \frac{\ket{\mathrm{vac}}_+ - \ket{\mathrm{vac}}_-}{\sqrt{2}}, \label{pbasis}
\end{align}
and write the Hamiltonian in the diagonal blocks structure $W \tilde{\mathcal{H}} W^\dagger = \tilde{\mathcal{H}}^+ \oplus \tilde{\mathcal{H}}^-$, with
\begin{eqnarray} 
    \tilde{\mathcal{H}}^+ &=& \begin{pmatrix}
        -\mu & \frac{\xi}{\sqrt{2}} & & \\
        \frac{\xi}{\sqrt{2}} & \mu + \frac{\pi}{3} & \frac{\xi}{2} & \\
         & \frac{\xi}{2} & -\mu + \frac{2\pi}{3} & \frac{\xi}{\sqrt{2}} \\
          & & \frac{\xi}{\sqrt{2}} & \mu + \frac{2\pi}{3}
    \end{pmatrix}, \nonumber\\
     \tilde{\mathcal{H}}^- &=& \begin{pmatrix}
        \mu + \frac{\pi}{3} & \frac{\xi}{2} \\
        \frac{\xi}{2} & -\mu + \frac{2\pi}{3}
    \end{pmatrix},
    \label{Eq::H+tilde}
\end{eqnarray}
where $W$ is the unitary operator rotating form the basis of Fig.~\ref{fig:phys_2sites} to the basis in Eqs.~\eqref{Eq::phys1}--\eqref{pbasis}

\begin{figure*}[t!]
    \centering
	\subfigure[]{\includegraphics[width=0.32\linewidth]{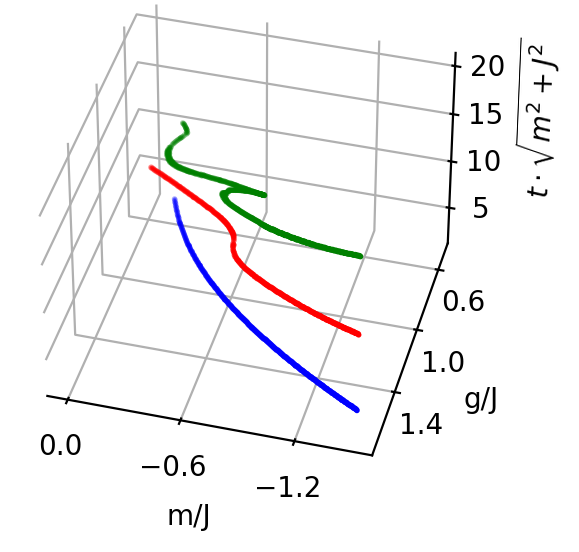}}
	\subfigure[]{\includegraphics[width=0.3\linewidth]{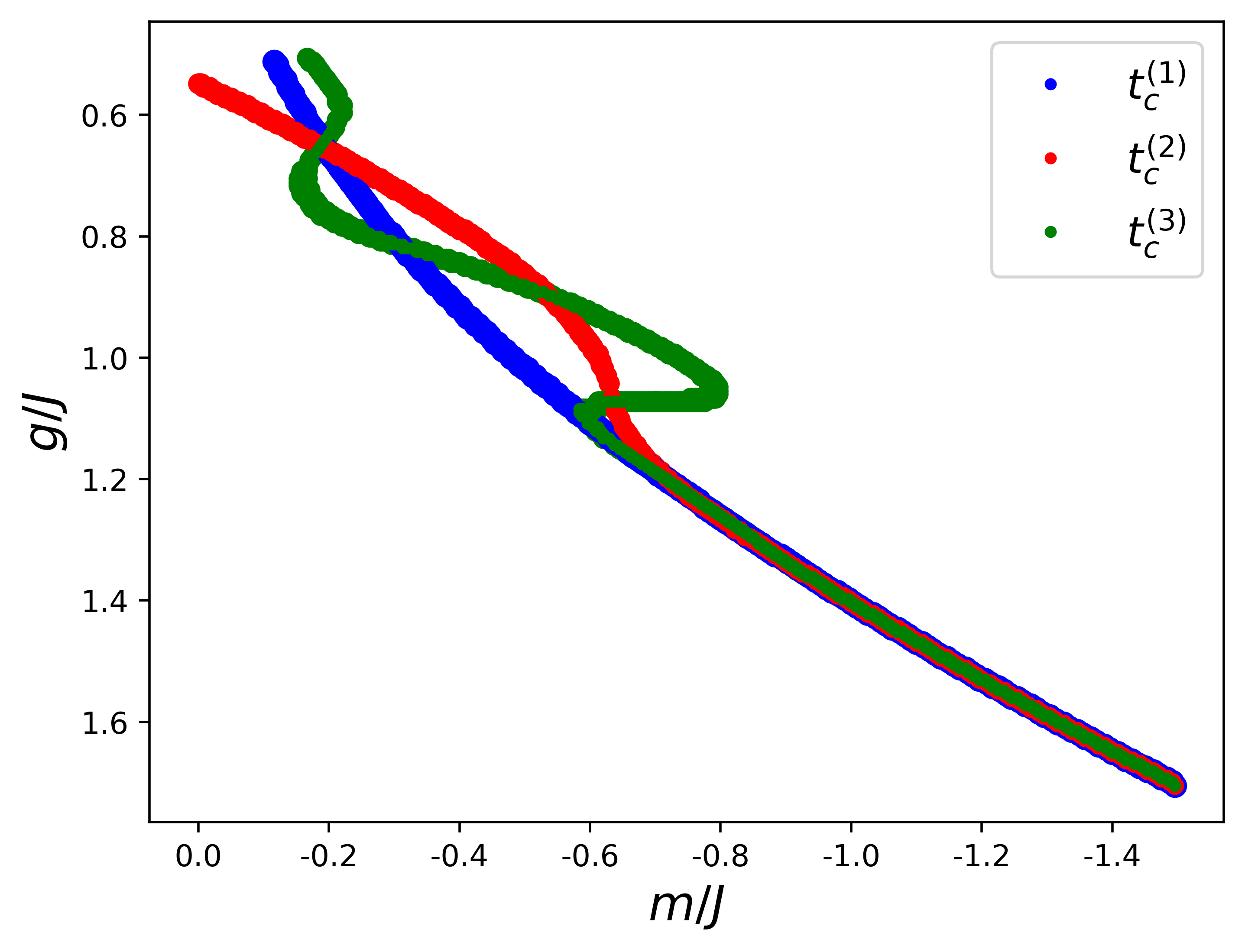}}
	\subfigure[]{\includegraphics[width=0.35\linewidth]{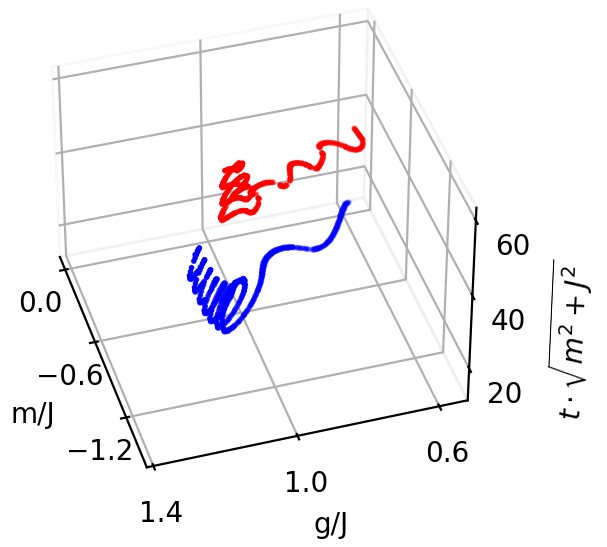}}
	
    \caption{Loschmidt amplitude zeros loci. {\bf Case 1.} In panel (a) a 3D plot representation of the parameters $(\frac{m}{J}, \frac{g}{J})$ of the quenched Hamiltonian $\mathcal{H}^+$ and the corresponding times $t\sqrt{m^2+J^2}$ required for the resonant Rabi oscillation of $\ket{\psi_1^+}$ and $\ket{\psi_2^+}$. The three colours correspond to the first three critical times. In panel (b) the projection of the 3D plot in (a) onto the $m-g$ plane is depicted. The zeros become periodic for each value of $(\frac{m}{J}, \frac{g}{J})$ below the critical mass $m_c \approx \frac{1}{\sqrt{2}}$. {\bf Case 2.} The same color code is exploited in panel (b) for the first two critical times of the resonant Rabi oscillation involving the initial Dirac vacuum $\ket{\psi_1^+},$ and a different mesonic state $\ket{\psi_4^+}$.}
    \label{fig:dqpt_traj}
\end{figure*}

\subsection{Dynamical quantum phase transition} \label{resonant_rabi}

For a general quench protocol, involving a system initially prepared in the ground state $\ket{\psi_0}$ of some Hamiltonian $\mathcal{H}_0$ and evolved with Hamiltonian $\mathcal{H}_1$, we can introduce the notion of Loschmidt (or survival) amplitude~\cite{dqpt,entropy}
\begin{equation}
    \mathcal{G}(t) = \bra{\psi_0} \e^{-\ii \mathcal{H}_1 t}\ket{\psi_0},
\end{equation}
which characterizes the deviation of the time-evolved state from the initial one.
The Loschmidt echo
\begin{equation}
    \mathcal{L}(t) = |\mathcal{G}(t)|^2 = \e^{-N \lambda(t)},
\end{equation}
describes  the probability of return to the initial state, and can be expressed in terms of the dynamical free energy (or rate function) $\lambda(t)$ and the number of degrees of freedom $N$, in analogy with the partition function and free energy of statistical mechanics.

DQPTs are defined as points in which the dynamical free energy becomes a non-analytic function of time, in the thermodynamic limit. Since for $N<\infty$ the Loschmidt echo is an analytic function of time, any non-analyticity in the rate function $\lambda(t)=-\frac{1}{N}\log(\mathcal{L}(t))$ comes from the zeros of $\mathcal{L}(t)$, in correspondence of which $\lambda(t)$ diverges. Thus possible DQPTs can be identified by finding these isolated zeros.

DQPTs have been thoroughly studied in the literature in recent years~\cite{dqpt}. In the case of 1D and 2D non-interacting systems, their phenomenology has been extensively understood in terms of the underlying equilibrium phase transition.
The simplest cases of study are two-band systems of non-interacting fermions that exhibit particle-hole symmetry. One can exploit translational invariance to write the Hamiltonian in a block diagonal form $\mathcal{H} = \bigoplus_{k}\mathcal{H}_k$, where each block Hamiltonian $\mathcal{H}_k$ acts non-trivially only on a two-dimensional subspace labeled by momentum $k$. In general, we have $\mathcal{H}_k = \textbf{c}_k^{\dag}h_k\textbf{c}_k$, with $\textbf{c}_k$ a spinor whose properties depend on the microscopic details of the Hamiltonian. For example, $\textbf{c}_k = (c_{ke}, c_{ko})$ for a chain of non-interacting staggered fermions, where the two operators $c_{ke}, c_{ko}$ act on the even/odd sites of the lattice.
Then the properties of the model are completely specified by the $2 \times 2$ Hermitian matrices $h_k$, which we can write in terms of the Pauli matrices as $h_k = \vec{d}_k \cdot \vec{\sigma}$, since the particle-hole symmetry of the Hamiltonian implies that the $h_k$ are traceless.

Any non-equilibrium protocol that does not couple the different momentum sectors can be expressed as the change in the vector field $\vec{d}_k(t)$, as shown in~\cite{dqpt_top}. For a quench protocol involving the ground state $\ket{\psi_0}$ of an initial Hamiltonian $\mathcal{H}_0$ with vector field $\vec{d}_{0k}$, evolved under the action of a Hamiltonian $\mathcal{H}_1$ with $\vec{d}_{1k}$, the Loschmidt amplitude can be exactly evaluated as
\begin{equation}
    \mathcal{G}(t) = \bra{\psi_0} \e^{-\ii\mathcal{H}_1t}\ket{\psi_0} = \prod_k \left[\cos(\omega_{1k}t) + \ii \frac{\vec{d}_{0k} \cdot \vec{d}_{1k}}{\vert\vec{d}_{0k}\vert \vert\vec{d}_{1k}\vert} \sin(\omega_{1k}t) \right],
\end{equation}
where $\omega_{1k}$ are the eigenvalues of $\mathcal{H}_1$. 
If $\vec{d}_{0k} \cdot \vec{d}_{1k} = 0$ for some value of $k$, a DQPT will appear at every critical time $t_c$, such that $\cos(\omega_{1k} t_c) = 0$, namely when $t_c = (n+1/2)\pi/\omega_{1k}$.

It is known in the literature~\cite{dqpt} that in 1D systems at least one such value of $k$ must satisfy this condition if, during the quench protocol, the Hamiltonian encounters an underlying equilibrium quantum phase transition from a trivial to a topological phase~\cite{zache2019}. The resulting DQPTs are labeled ``topological", since they appear independently from the choice of the parameters of the Hamiltonians $\mathcal{H}_0$ and $\mathcal{H}_1$, as long as they are characterized by different winding numbers. A DQPT might be present even in the absence of an underlying transition, labeled in this case as ``accidental", whose realization requires a fine-tuning of the Hamiltonian.

These properties are well defined in the thermodynamic limit. However, zeros of the Loschmidt amplitude, signalling possible DQPTs, can be identified already for very small lattice sizes. We show as an example the 2-site system characterized by the Hamiltonian in Eq.~\eqref{Eq::Hximu}, with antiperiodic boundary conditions. The advantage of having small-size systems is that their dynamics can be simulated with high fidelity on a realistic noisy quantum simulator.

In our problem, in the non-interacting case $g=0$, the Hamiltonian in Eq.~\eqref{Eq::HmJ} is characterized by a vector field $\vec{d}_k = \vec{d} = (0, -J, m)$ with eigenvalues $\pm\omega = \pm\sqrt{m^2 + J^2}$. A quench induced by a change in the mass on the ground state of $\mathcal{H}_0=\mathcal{H}(\bar{m},J,g=0)$ with Hamiltonian $\mathcal{H}_1=\mathcal{H}(m,J,g=0)$ yields a Loschmidt amplitude
\begin{equation}
\mathcal{G}(t) = \cos(\omega t) + \ii \frac{\bar{m}m+J^2}{\bar{\omega}\omega}\sin(\omega t),  
\end{equation}
meaning that a possible DQPT arises when $\bar{m}m+J^2=0$ at time scales of order $t \sqrt{m^2+J^2} \sim 1$.
A case of interest is when the initial ground state is the Dirac vacuum, i.e. the ground state of $\mathcal{H}(\bar{m},J=0,g=0)$, in which the Loschmidt amplitude vanishes only if $m=0$.


In the interacting case $g\neq 0$, even for $N=2$ lattice sites an analytical solution is lacking, and we need to resort to numerical diagonalization for different values of the parameters $m, J$. We consider again a quench protocol involving an initial Dirac vacuum, with a sudden variation to finite mass. Physical subspace sectors $\mathscr{H}_G = \mathscr{H}_G^+ \oplus \mathscr{H}_G^-$ are exploited in the encoding because for the initial Dirac vacuum $\ket{\psi_1^+} \in \mathscr{H}_G^+$
\begin{equation}
    \ket{\psi_1^+(t)} = U(t) \ket{\psi_1^+} = \mathrm{e}^{-\ii \mathcal{H}^+ t} \ket{\psi_1^+} \in \mathscr{H}_G^+,
\end{equation}
thus limiting noise affecting the implemented dynamics within the selected sector.

\begin{figure*}[t!]
    \centering
    \includegraphics[width=1\textwidth]{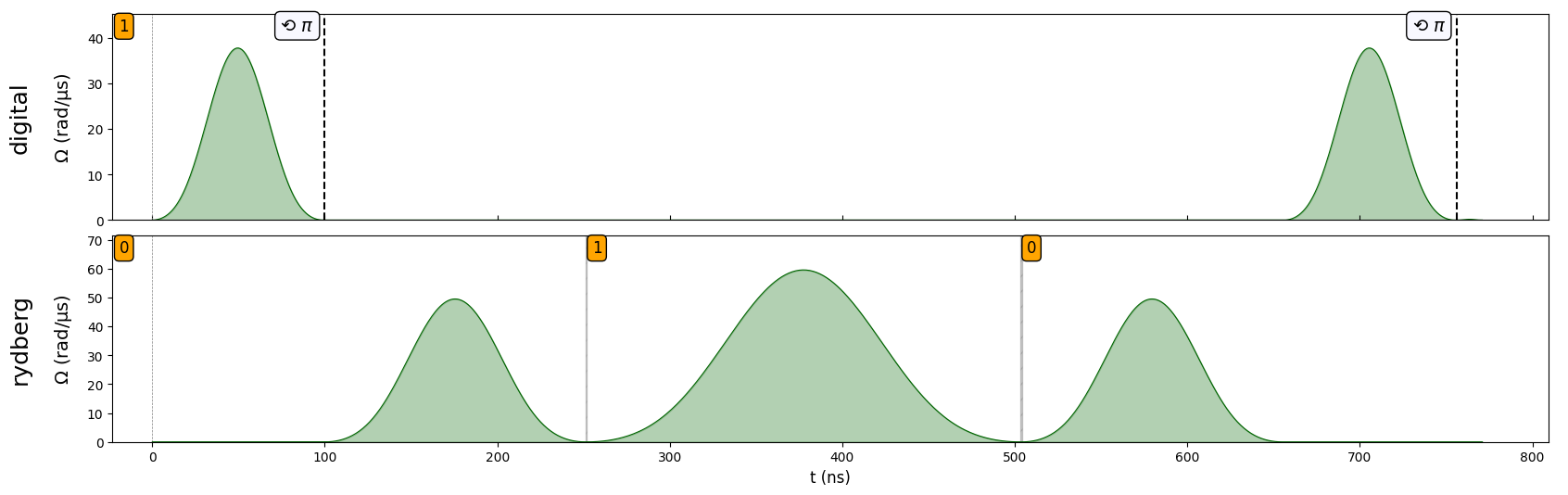}
    \caption{Pulse sequence to implement the CNOT gate by using a CZ and two Hadamard gates, represented in the digital channel by a rotation generated by $Y$ followed by a virtual $Z$ depicted as a vertical dashed line.}
    \label{fig:cnot pulses}
\end{figure*}

The results of our numerical simulations are shown in Fig.~\ref{fig:dqpt_traj}, where we plot the loci of the vanishing Loschmidt amplitude (or, equivalently, a diverging rate function $\lambda (t)$) as function of the parameters $(m/J, g/J)$ of the quenched Hamiltonian and the times $t \sqrt{m^2+J^2}$, rescaled by a factor $\sqrt{m^2+J^2}$ in analogy with the non-interacting case. We examine, in particular, two cases.

{\bf Case 1.} Fig.~\ref{fig:dqpt_traj} (a) shows the zeros of Loschmidt amplitude for the first three times.  Interestingly, below a critical value of the mass parameter $m_c/J\lesssim -1/\sqrt{2}$, as highlighted in Fig.~\ref{fig:dqpt_traj} (b), we observe a superposition of all critical times signalling a periodicity of the DQPTs for each $(m/J, g/J)$ pair. To understand this phenomenology, we look at the form of the Hamiltonian corresponding to these particular parameters. First, we notice that the zeros of the Loschmidt amplitude below the critical value corresponds to a functional relation between $g$ and $m$, which we investigated numerically by means of linear regression finding  $\log(g/J) \approx 0.335+0.49 \log(-m/J)$. This result is in agreement with the analytic relation $g/J =  A\sqrt{-m/J}$ with $A = \sqrt{\frac{6}{\pi}}$, that can be deduced from Eq.~\eqref{Eq::H+tilde} by considering the expression of  $\tilde{\mathcal{H}}^+$ and imposing the equality between the first and the second diagonal element. Indeed, in this case we have that the Hamiltonian
\begin{equation} \label{Eq::rabi}
    \mathcal{H}^+ \equiv \frac{g^2}{J}\tilde{\mathcal{H}}^+  = \begin{pmatrix} 
        0 & \frac{J}{\sqrt{2}} & & \\
        \frac{J}{\sqrt{2}} & 0  & \frac{J}{2} & \\
         & \frac{J}{2} & 2\Delta & \frac{J}{\sqrt{2}} \\
          & & \frac{J}{\sqrt{2}} & \Delta
    \end{pmatrix} - m\mathbf{1},
\end{equation}
is that of a 4-level system with Rabi transitions where the last two states $\ket{\psi_3^+}, \ket{\psi_4^+}$ are detuned with respect to resonant $\ket{\psi_1^+}, \ket{\psi_2^+}$ by a factor $\Delta = -2m$. We thus observe a phenomenology corresponding to what is discussed in~\cite{dqpt_QLM,dqpt_QLM2}, with the initial Dirac vacuum $\ket{\psi_1^+}$ evolving only in the subspace spanned by itself and the mesonic ground state $\ket{\psi_2^+}$. The periodicity in the DQPTs then corresponds to the periodicity of Rabi oscillations. The approximation of a resonant Rabi model improves with decreasing values of $m/J$, meaning that the population of the detuned states is small but non-zero near $m_c$, and it decreases towards a vanishing value for $m<m_c$.

{\bf Case 2.} In general the form of the Hamiltonian in Eq.~\eqref{Eq::H+tilde} imposes that every time two diagonal elements have the same value, the corresponding states will exhibit this kind of resonance. Considering once again the case in which the initial state is the Dirac vacuum, we are interested in values of $(m/J, g/J)$ for which the first diagonal element is equal to one of the last two. A quick calculation shows that this happens when $g/J = \frac{A}{\sqrt{2}}\sqrt{-m/J}$, such that the first and the last diagonal elements are equal, which implies a resonant Rabi model between $\ket{\psi_1^+}$ and $\ket{\psi_4^+}$. This family of Hamiltonians keeps the same off-diagonal structure of Eq.~\eqref{Eq::rabi}, while the diagonal part reads $\mathrm{diag}\{ \mathcal{H}^+ \} = (-m, 0, -3 m, -m)$. On the contrary, there is no value of $g\neq 0$ for which the third diagonal element equals the first. The observed DQPTs require longer simulation times $t \sqrt{m^2+J^2}$, with a longer period of oscillations as shown in Fig.~\ref{fig:dqpt_traj} (c), because the involved states are not directly coupled by the Hamiltonian.

\section{Simulation on Rydberg platform} \label{simulation}

\subsection{Pulse sequence}

Neutral atom quantum processors (QPU) have the capability to execute both digital and analog quantum processing tasks~\cite{briegel2000quantum}. In the realm of digital computing, quantum algorithms are deconstructed into a sequence of quantum logic gates. These gates are implemented by precisely directing laser pulses onto specific atoms within the register. The state of each individual qubit is then measured to probe the final state of the system. This section focuses on the implementation of digital computing through pulses sequence engineered in order to perform the time evolution of our model. 

Atoms in the Rydberg states exhibit an extraordinary electric dipole moment, which can be three orders of magnitude larger than in their ground state. Consequently, when two Rydberg atoms are positioned at a short distance of a few micrometers, they experience a strong dipole-dipole interaction capable of significantly affecting the energy of the doubly excited state. This interaction effectively prevents the simultaneous excitation of both atoms. This phenomenon, known as Rydberg blockade, forms the essential mechanism for realizing quantum logic. Neutral atom devices can be summarized as having two main components.
\begin{itemize}
    \item[a)] \textbf{The register.} This component consists of a group of trapped atoms arranged in a suitable configuration. Each atom possesses a specific quantum state encoded in distinct electronic levels.
    \item[b)] \textbf{The channels.} The manipulation of the state of the atoms is entrusted to these components, as they address specific electronic transitions.
\end{itemize}
At the core of the system is the sequence of pulses. This sequence, along with other instructions, is sequentially assigned to the channels. Each pulse represents a modulation of a channel's output amplitude, detuning, and phase over a specific duration. While the phase remains constant within a pulse, the amplitude and detuning are determined by waveforms. These waveforms dictate the amplitude and detuning values throughout the pulse's duration. We will exploit Blackman waveforms in order to minimize photon leakage~\cite{pulser}.

By utilizing a control field, it becomes possible to execute arbitrary rotations of the qubit state on the Bloch sphere. In these devices, this control field is represented by an optical laser field that drives electronic transitions through an intermediate atomic state~\cite{PhysRevLett.96.063001, levine2019parallel}. The interaction between the atom and the laser is characterized by various factors: the Rabi frequency $\Omega$ (which determines the strength of the interaction and it is proportional to the amplitude of the laser field), the detuning $\delta$ (which denotes the discrepancy between the qubit resonance and the field frequencies), and their relative phase $\phi$.

In this system, a pulse acting on an atom at site $i$ with Rabi frequency $\Omega(t)$, detuning $\delta(t)$ and a fixed phase $\phi$, will have the Hamiltonian terms
\begin{equation}
    \frac{\hbar \Omega(t)}{2} \bigl(X_i \cos(\phi) - Y_i \sin(\phi)\bigr) -\frac{\hbar}{2} \delta(t) Z_i .
\end{equation}
Once a certain axis with respect to $\phi$ is chosen, the rotation angle related with a pulse duration $\tau$ is
\begin{equation}
    \theta = \int_0^\tau \Omega(t) \mathrm{d} t.
\end{equation}
As a result, adjusting the pulse duration, laser intensity, detuning, and phase of the laser enables the implementation of any single-qubit gate.

\begin{figure*}[t!]
    \centering
    \subfigure[]{\includegraphics[width=1\textwidth]{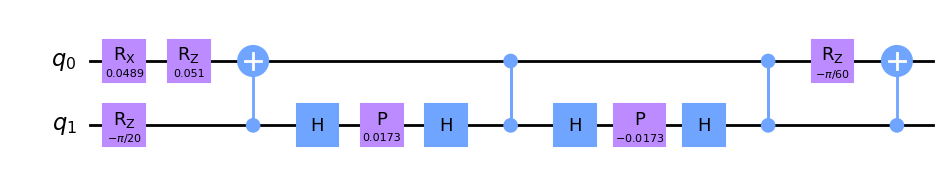}}
    \subfigure[]{\includegraphics[width=1\textwidth]{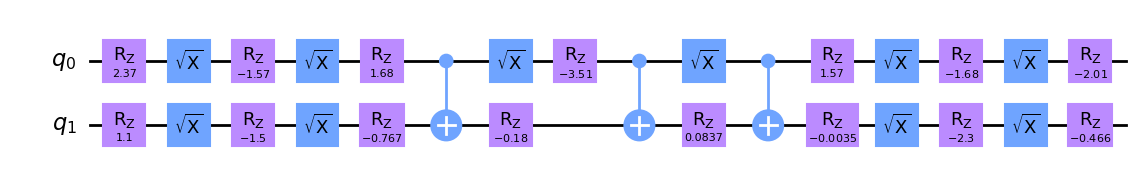}}
    \caption{(a) Circuit implementation of one time step in Trotter's formula for the evolution operator $e^{-\ii \mathcal{H}^+ t}$ approximation. (b) The resulting circuit after applying error suppression technique by Qiskit for the composition of $91$ time steps. We choose here the following parameters: $\xi \approx 0.346$ and $\mu \approx -0.517$.}
    \label{fig:trotter_step}
\end{figure*}

To achieve a universal quantum gate set, all that is required is an appropriate two-qubit gate, alongside the ability to perform arbitrary single-qubit gates. In neutral-atom devices, the controlled-Z (CZ) gate serves as the native gate. Its implementation capitalizes on the Rydberg blockade effect to conditionally flip the phase of a two-qubit subsystem based on its initial state. This is accomplished by sequentially bringing the atoms to an auxiliary Rydberg state and then back, utilizing a sequence of three consecutive laser pulses as described in~\cite{pulser}. This means that the space in which the atom arrays can evolve consists of three states: two non-interacting hyperfine states, denoted as $\ket{g}$ and $\ket{h}$, and an auxiliary Rydberg state, labeled as $\ket{r}$. The auxiliary Rydberg state $\ket{r}$ is exclusively populated during the transient dynamics of the system while multi-qubit gates are being applied. By incorporating Hadamard gates on the target atom both before and after the operation, the CZ gate can be utilized to produce the CNOT gate: the resulting pulse sequence is shown in Fig.~\ref{fig:cnot pulses}. The maximization of the Rydberg blockade is referred in our simulated setup to a minimal available distance between the two neutral atoms equal to $4 \mu$m. Indeed, knowing that the Rydberg radius as a function of the Rabi frequency is given by
\begin{equation}
R_b = \left( \frac{C_6}{\hbar \Omega} \right)^{\frac{1}{6}},
\end{equation}
with $C_6$ being a constant, a distance of $4 \,\mu$m ensures that the system remains in the blockade regime for the entire range of $\Omega$ values of the pulses, whose upper bound is approximately equal to $62.831 \,\mathrm{rad}/\mu$s for the chosen $\texttt{Chadoq2}$ device.

\begin{figure*}[t!]
    \centering
    \includegraphics[width=1\textwidth]{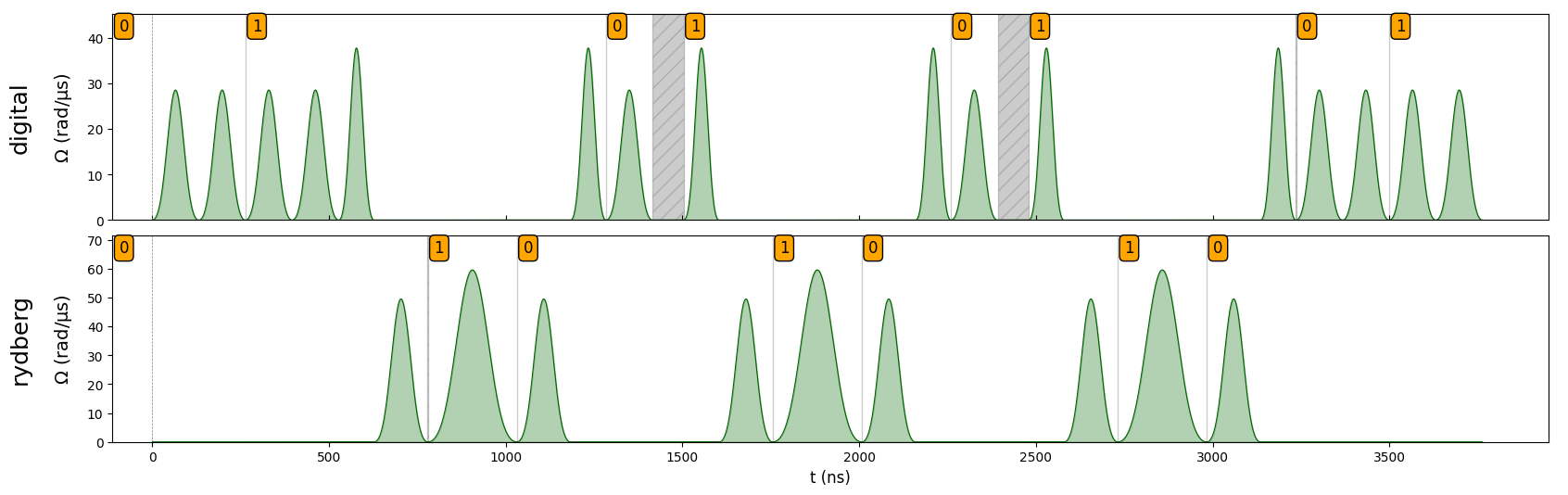}
    \caption{Pulse sequence needed to implement the optimized circuit shown in Fig.~\ref{fig:trotter_step} (b). Each pulse in the digital channel is followed by a virtual $Z$ phase shift.}
    \label{fig:pulser_trotter_step}
\end{figure*}

\subsection{Trotter evolution}

Parity conservation during the evolution allows us to encode just the sector of the initial Dirac vacuum state. It is spanned by states in Eq.~\eqref{Eq::phys1}--\eqref{Eq::phys2}, thus requiring a system composed by two qubits. Any observable is expressed in terms of tensor products of Pauli matrices $\{ G_{k \ell} \} = \{P_k \otimes P_\ell /2\}_{k,\ell = 0,1,2,3}$, where $P_k \in \{ \mathbbm{1}, X, Y, Z \}$,
\begin{equation}
    \mathcal{O} = \sum_{k,\ell=0}^3 \mathrm{Tr} \{ G_{k \ell}\mathcal{O} \} G_{k \ell},
\end{equation}
In particular, the considered diagonal block Hamiltonian $\tilde{\mathcal{H}}^+$ in~\eqref{Eq::H+tilde} is decomposed as
\begin{align} \label{eq:decomposition}
\tilde{\mathcal{H}}^+=& \ \frac{\xi}{4}\Bigl( X \otimes X + Y \otimes Y + 2\sqrt{2}(\mathbbm{1} \otimes X)\Bigr) \nonumber\\
&-\frac{\pi}{12}\Bigl(Z \otimes Z + 3 (Z \otimes \mathbbm{1})\Bigr) \\ \nonumber &-\bigg(\frac{\pi}{12}+\mu \bigg) (\mathbbm{1} \otimes Z ) + \frac{5\pi}{12} (\mathbbm{1} \otimes \mathbbm{1}),
\end{align}
where the last term will be safely neglected because it yields just an overall phase for time evolution.

The evolution operator $\e^{-\ii \mathcal{H}^+ t}$ is implemented by Trotter's formula. At each time step, the various terms in Eq.~\eqref{eq:decomposition} can be implemented by
\begin{align}
&\mathrm{e}^{-\ii\frac{\xi t}{4}\big( X \otimes X + Y \otimes Y \big) +\ii\frac{\pi t}{12} \big( Z \otimes Z  \big)} = \ \text{CNOT}_{01}\text{H}_{1}\text{P}_{1}\Bigl(\frac{\xi t}{2}\Bigr)\,  \text{H}_{1} \text{CZ}_{01} \nonumber\\
& \qquad\qquad \times\text{H}_{1}\text{P}_{1}\Bigl(-\frac{\xi t}{2}\Bigr)\,\text{H}_{1} \text{CZ}_{01} R^Z_0\Bigl(-\frac{\pi t}{6}\Bigr)\, \text{CNOT}_{01} ,
\end{align}
where $0$ and $1$ refer to the target and control qubits, H and P are the Hadamard and phase shift gates respectively. The remaining terms of $e^{-\ii \mathcal{H}^+ t}$ are just $R^X$ and $R^Z$ quibit rotations:
\begin{align}
&\mathrm{e}^{-\ii t\frac{\sqrt{2}}{2}\xi  (\mathbbm{1} \otimes X)}
\mathrm{e}^{\ii t(\frac{\pi}{12} +\mu) (\mathbbm{1} \otimes Z)+\ii t\frac{\pi}{4} ( Z \otimes \mathbbm{1})} \nonumber\\
& \qquad\qquad=  \ R_0^Z\Bigl(-\frac{\pi t}{4}\Bigr)\,R^X_0\Bigl(\frac{\sqrt{2} \xi t}{2}\Bigr)  \, R^Z_1\biggl(\Bigl(-\mu +\frac{\pi}{12}\Bigr)t\biggr),
\end{align}
where $R_j$ stands for a rotation of qubit $q_j$.

The final circuit for a single Trotter step is shown in Fig.~\ref{fig:trotter_step}(a).
The composition of multiple time steps evolution can be further optimized by exploiting error suppression techniques of \texttt{qiskit}~\cite{Qiskit} that transform your compiled circuit in order to minimize depth and consequently errors. We set the optimization level equal to $3$: the resulting circuit and its implementation in a pulse sequence can be seen in Fig.~\ref{fig:trotter_step}(b) and Fig.~\ref{fig:pulser_trotter_step} respectively. The stable structure of the optimized circuit for increasing number of Trotter steps is motivated by the decomposition of any two qubits gate with three CNOT gates proved in~\cite{3cnot}.

\subsection{Noise models} \label{noise_mod}

The implementation of the quantum gates on real quantum device is subject to errors and noise. A detailed explanation of these phenomena in the Rydberg platforms can be found in~\cite{de2018analysis}. A simulation of a pulse sequence is able to reproduce them accurately as well. Here we briefly describe the ones we are going to take into account in this work.

\begin{itemize}
    \item[a)] \textbf{State preparation errors} arise when an atom fails to be correctly prepared in the ground state at the beginning of the process, with a probability denoted by $\eta$. As a result, the atom remains effectively unavailable throughout the sequence.
    \item[b)]\textbf{Measurement errors} occur due to the inaccurate identification of the final state of the atoms. During the imaging process, some excited Rydberg atoms in the state $\ket{r}$ might decay to the ground state $\ket{g}$, causing them to be mistakenly trapped in the tweezers. This leads to false negatives, which occur at a rate denoted by $\epsilon$. Conversely, some atoms in the ground state $\ket{g}$ might be unintentionally expelled from the system due to various factors, such as collisions with residual gas in the chamber. Additionally, the recapture of these atoms by the tweezers may fail, resulting in incorrect identification as atoms in the excited state $\ket{r}$. These errors are known as false positives and occur at a rate denoted by~$\epsilon'$.
    
    In summary, State Preparation and Measurement (SPAM) errors encompass state preparation errors, measurement errors leading to false negatives, and false positives arising from the misidentification of atoms. 
    \item[c)] \textbf{Doppler damping.} The atoms within the register are cooled to a low, yet non-zero temperature $T\sim 50\mu$K. Consequently, the laser frequency they perceive undergoes a Doppler shift due to their thermal motion. This shift manifests as a detuning frequency alteration in the laser.
    \item[d)]\textbf{Amplitude fluctuations.} A real laser amplitude exhibits fluctuations between pulses. These fluctuations can result in variations in the laser's intensity from one pulse to the next.
\end{itemize}

The translation of an optimized circuit into a pulses sequence requires a careful minimization of the overall duration, in order to suppress some of the noise sources. Any rotation generated by $Z$ is time demanding, since e.g. $R^Z(\theta) = R^X(\frac{\pi}{2})\, R^Y(\theta)\, R^X(-\frac{\pi}{2})$, thus involving a too high number of pulses. Instead the adoption of virtual Z gates completely solves this issue by means of instantaneous phase shifts~\cite{virtualZ}. In Fig.~\ref{fig:pulser_trotter_step} it is possible to recognize the CNOT sequences shown in Fig.~\ref{fig:cnot pulses}, related with the three CNOTs provided by \texttt{qiskit} optimization in Fig.~\ref{fig:trotter_step} (b), but without an explicit representation of virtual $Z$ for clarity. Indeed the high number of $R^Z$ in Fig.~\ref{fig:trotter_step} (b) yields a virtual $Z$ after each pulse in the digital channel. The overall duration of each circuit is almost equally split between CNOTs and $\sqrt{X}$ gates. Errors intrinsically related with the setup, like SPAM, cannot be tailored by duration minimization and they will result with a high contribution.

\section{Results}\label{results}

The simulations of experimental Rydberg atoms setups are available in \texttt{pulser}~\cite{pulser} by means of the tools introduced in Section~\ref{noise_mod}. Our test of this neutral atom platform is referred to a Schwinger model consisting in a periodic lattice composed by $N=2$ sites and an equal number of links endowed with a $\mathbb{Z}_3$ discretization of the $U(1)$ gauge group~\cite{Zn}. As discussed in Section~\ref{model}, we exploit the conservation law associated with parity in order to encode the minimal number of degrees of freedom: this strategy allows us to benefit from \texttt{qiskit} optimization~\cite{qiskit_opt} of two-qubits circuits, yielding a fixed depth for any long-time Trotter evolution, as shown in Fig.~\ref{fig:trotter_step}. The qubit pair is mapped into two neutral atoms separated by a distance equal to $4  \mu$m.

\begin{figure}[t!]
	\subfigure[]{\includegraphics[width=0.47\linewidth]{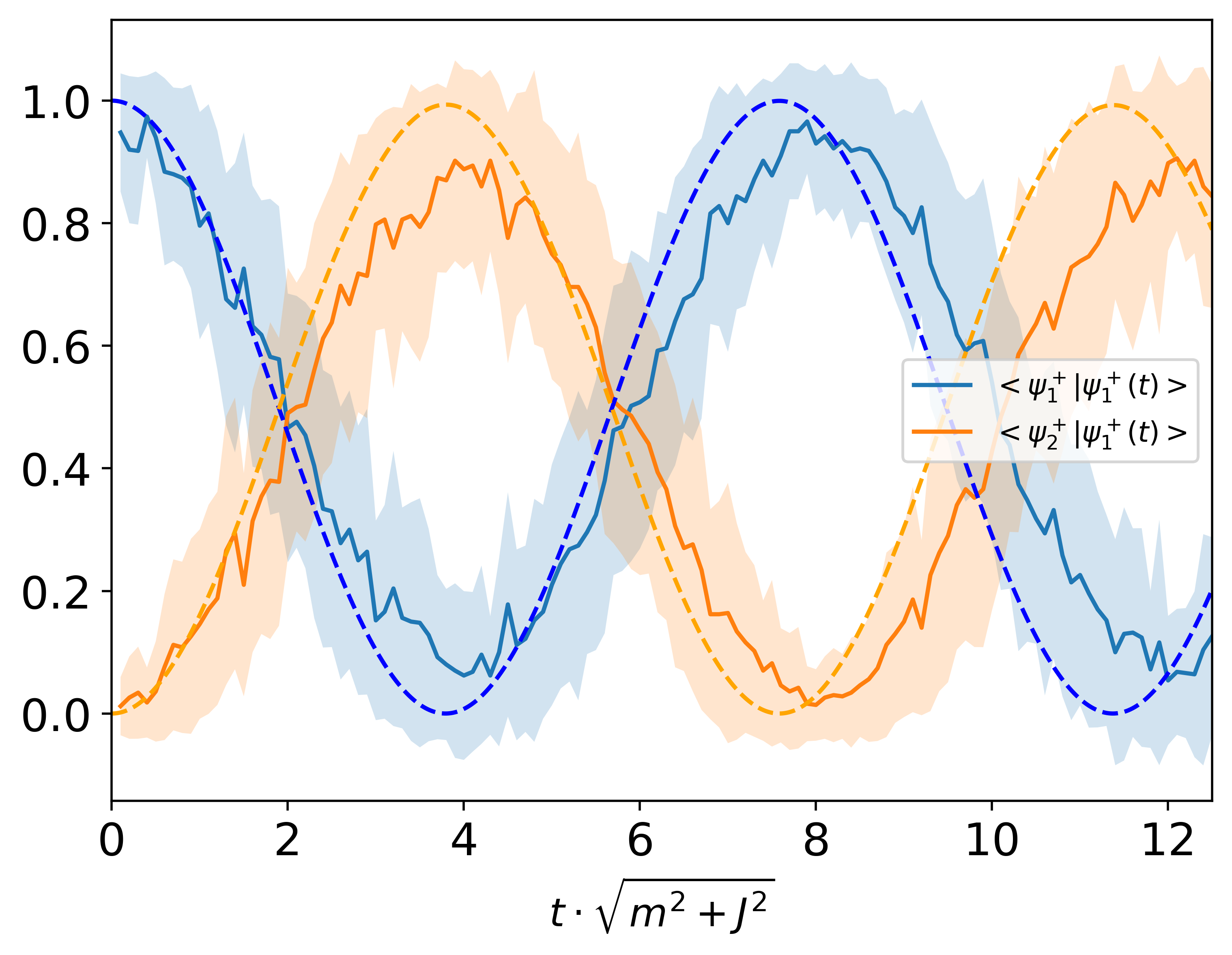}}
	\subfigure[]{\includegraphics[width=0.48\linewidth]{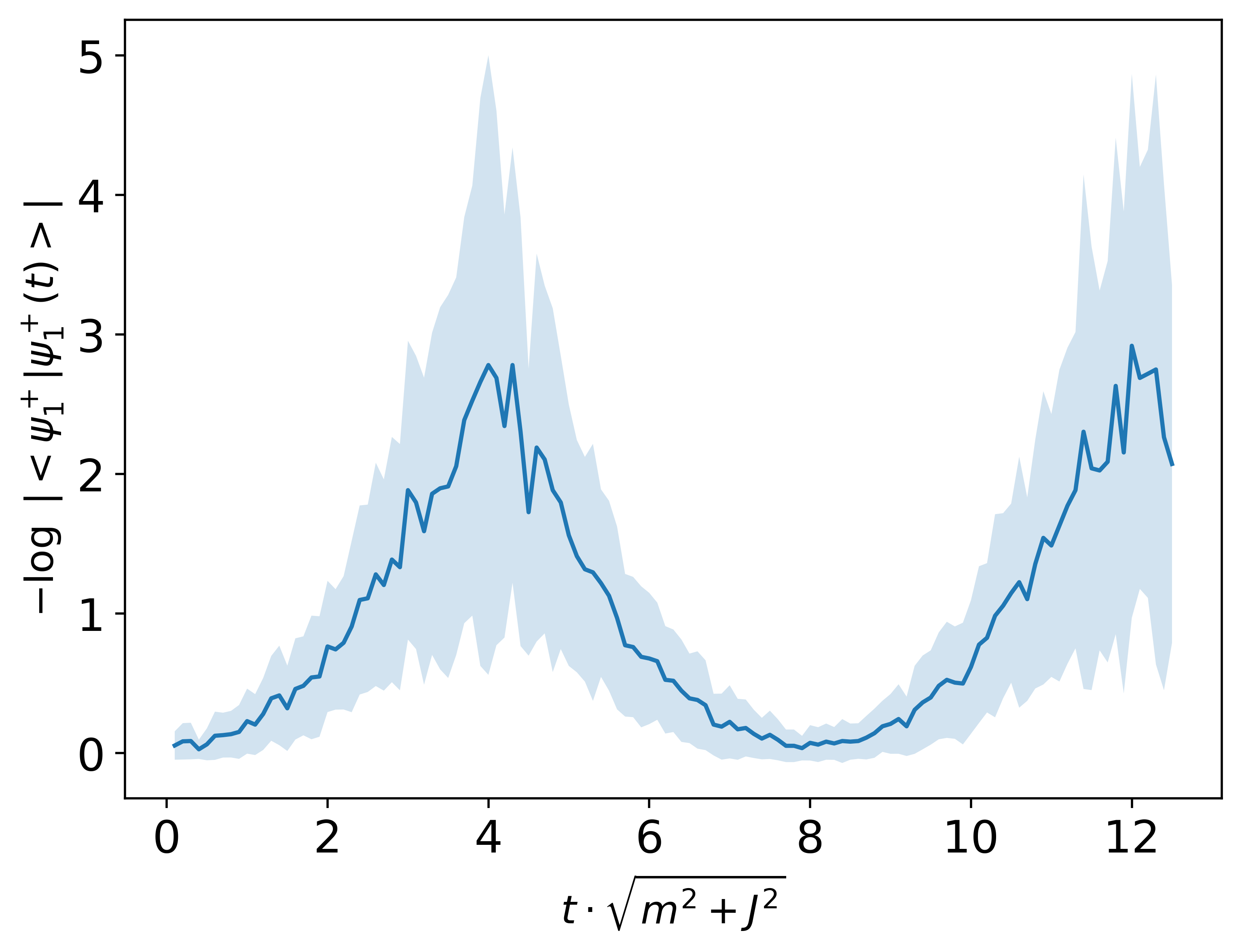}}\\
	\subfigure[]{\includegraphics[width=0.47\linewidth]{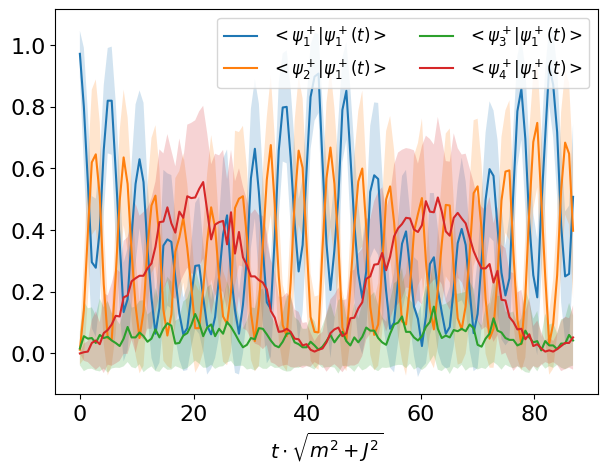}}
    \subfigure[]{\includegraphics[width=0.48\linewidth]{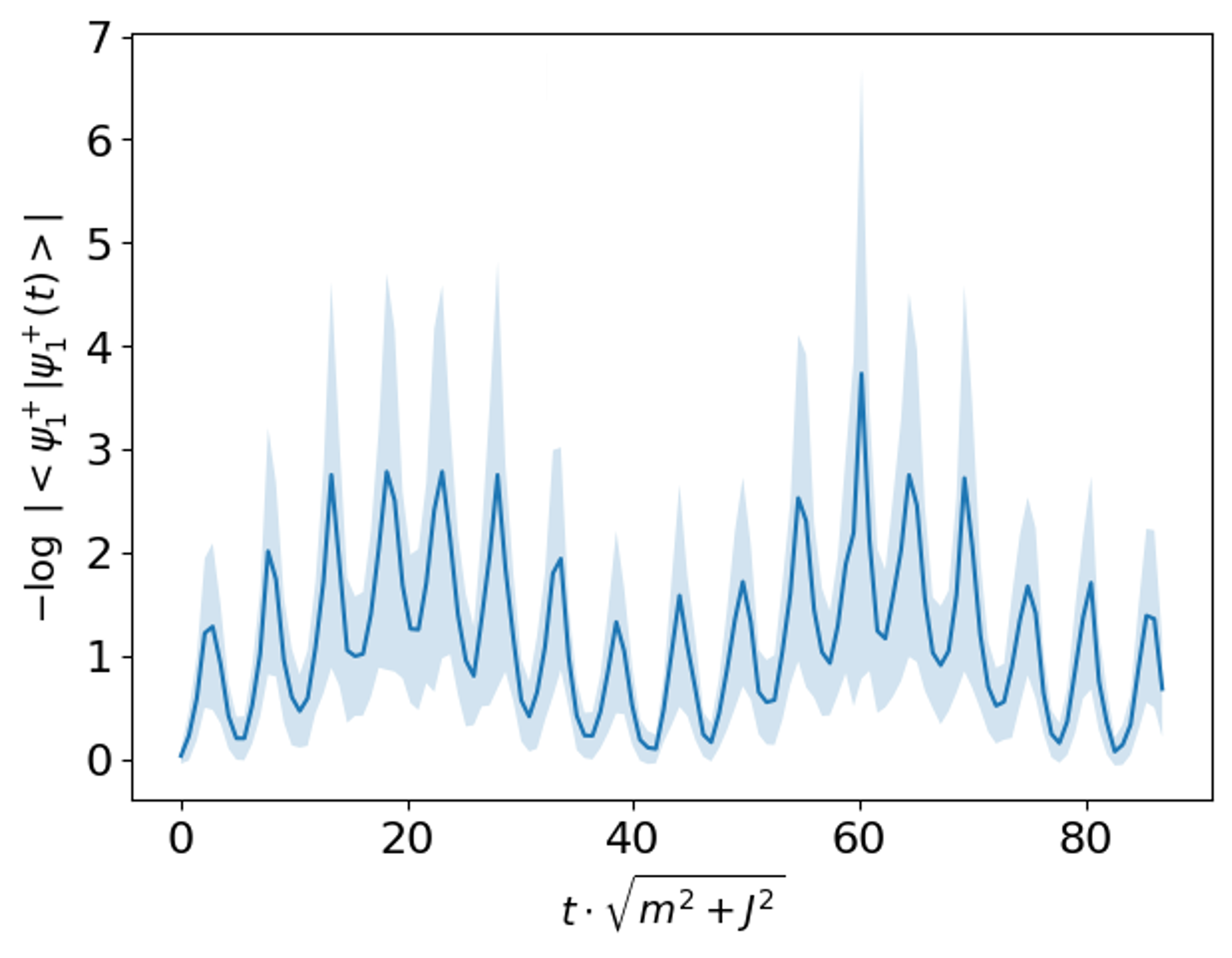}}
    \caption{Rabi oscillations between Dirac vacuum and mesonic states. Case 1 of  Eq.~\eqref{Eq::H+tilde} corresponding to Rabi oscillations between two states is analysed in panel (a), where the populations of the two states versus time are shown for $m/J\approx - 1.49$, $g/J = 1.7$ with dashed lines referring to noiseless evolutions and in panel (b) where we plot the rate functions versus time
    for the same values of parameters. Similarly, panels (c) and (d)   show the Rabi oscillations and rate function of Case 2  of  Eq.~\eqref{Eq::H+tilde} with $m/J\approx - 0.87$, $g/J = 1.05$. In panel (c) we omit for clarity noiseless evolutions, which nonetheless lie within one standard deviation.} \label{fig:rabi_dqpt}
\end{figure}

Our main purpose is the observation of multiple DQPTs corresponding to the resonant Rabi oscillations presented in Section~\ref{resonant_rabi}.
We  first consider a condition in the parameter space sufficiently below the threshold $m_c$ required for multiple DQPTs, namely the smallest mass $m/J \approx -1.49$ and highest charge $g/J = 1.7$ of the curves in Fig.~\ref{fig:dqpt_traj} (b). In this case both $\ket{\psi_3^+}$ and $\ket{\psi_4^+}$ are highly detuned, so that states population is concentrated in the subspace spanned by the resonant Dirac vacuum and the mesonic states. The \texttt{pulser} simulation includes the combination of each noise source introduced in Section~\ref{noise_mod} and the pulse sequence protocol described therein. We implemented $125$ Trotter steps in order to observe two DQPTs, but the simulated dynamics can be extended for an arbitrarily long time since the optimized circuit in Fig.~\ref{fig:trotter_step} (b) has fixed depth for any number of Trotter steps. As instance of Case 1 DQPT as discussed in Sect.~\ref{resonant_rabi}, we choose $m/J\approx - 1.49$, $g/J = 1.7$: in Fig.~\ref{fig:rabi_dqpt} (a) we show the Loschmidt echo averaged over $100$ noise realizations, as function of the rescaled time step $\Delta t  \sqrt{m^2 + J^2} = 0.1$, where the shading region gives one standard deviation at least when considering the first two periods. We observe Rabi oscillations period equal to $8$, which include  the noiseless evolution within one standard deviation. The correct location of rate functions divergences is also verified in Fig.~\ref{fig:rabi_dqpt} (b) for different coupling values increasing towards the threshold $m_c$, showing equal maxima for any averaged evolution, thus reaching the same minimal values as in Fig.~\ref{fig:rabi_dqpt} (a). 

\begin{figure}[t!]
	\includegraphics[width=0.95\linewidth]{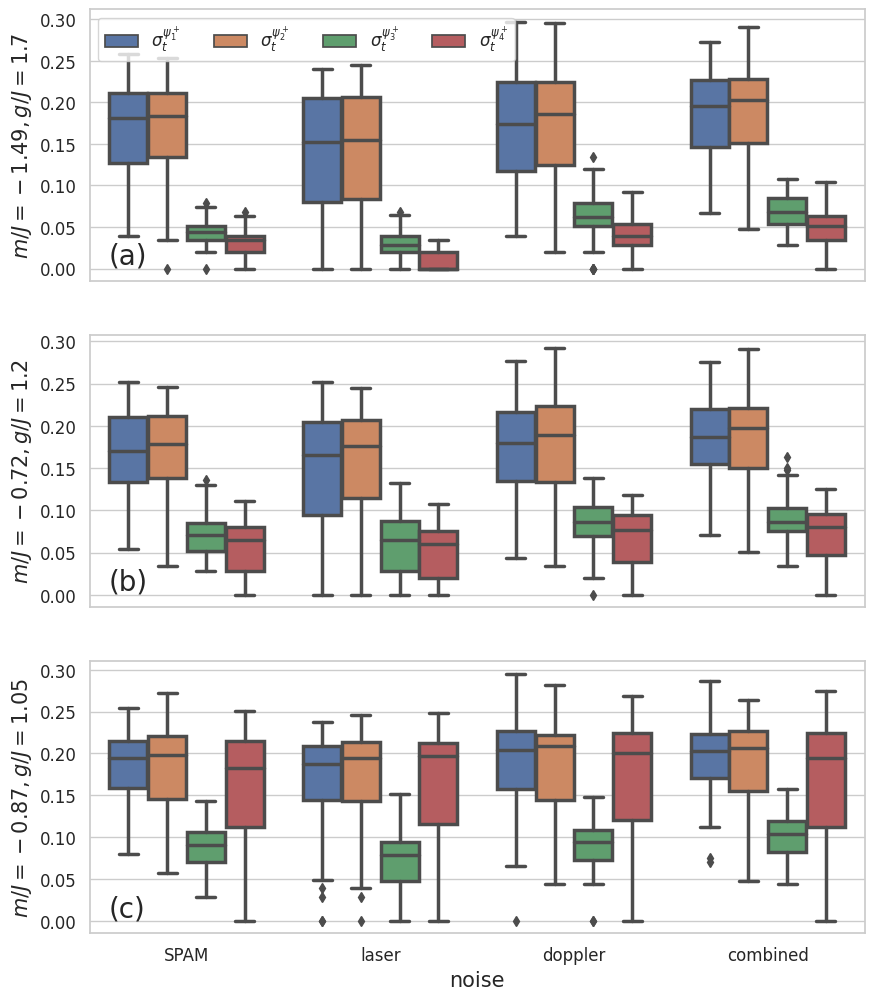}
    \caption{Fluctuations statistics collected over $100$ noise realizations of each noise model for three different parameters regimes. The top boxplot is referred to high detuning, as  in Fig.~\ref{fig:rabi_dqpt} (a), the middle one refers to a condition in proximity of $m_c$, while the bottom case includes just one DQPT well above $m_c$.} \label{fig:boxplot}
\end{figure}

In a region of parameter space close to $m_c$ from below, the average population of $\ket{\psi_3^+}$ and $\ket{\psi_4^+}$ over a resonant Rabi oscillation period is slightly above zero ($\approx 2 \times 10^{-2}$), because the detuning is not sufficient to achieve a complete population concentration in the subspace spanned by $\ket{\psi_1^+}$ and $\ket{\psi_2^+}$. The aforementioned non-negligible populations are detected in terms of the standard deviation statistics of each amplitude 
\begin{equation}
  \sigma^{\psi_i^+}_t = \sqrt{\mathbb{V}\left[\left|\braket{\psi_i^+}{\psi_1^+(t)}\right|^2\right]}
\end{equation}
determined with respect to the $100$ noise realizations. For example, cases characterized by high detuning with respect to resonant states show the same statistical fluctuation of $m/J=-1.49$ and $g/J=1.7$ as shown in the first panel of Fig.~\ref{fig:boxplot}. The median standard deviation of the laser amplitude fluctuation model is lower than the ones of the SPAM and Doppler models, which are approximately equal. In the combined model case, this median standard deviation is further increased. An interquartile range approximately equal to $0.1$ is observed for any noise model corresponding to $\sigma^{\psi_1^+}_t$ and $\sigma^{\psi_2^+}_t$, while negligible fluctuations characterize detuned states. On the contrary, in the proximity of $m_c$, we find a much higher variation of $\sigma^{\psi_3^+}_t$ and $\sigma^{\psi_4^+}_t$ as shown in the middle panel of Fig.~\ref{fig:boxplot}, due to the aforementioned non-negligible population. In this case, the noise model statistics are almost indistinguishable, except for lower median values for the off-resonance states in the case of SPAM and laser models.

The results for Case 2 of Section~\ref{resonant_rabi} corresponding to the resonant Rabi oscillation involving $\ket{\psi_4^+}$ through an indirect coupling are given in Fig.~\ref{fig:dqpt_traj} (c), in order to investigate richer dynamics still involving multiple DQPTs. The Rydberg simulation shown in Fig.~\ref{fig:rabi_dqpt} (c), corresponding to $m/J \approx -0.87$ and $g/J \approx 1.05$,  matches the exact evolution of each state within the shaded region, quantifying the standard deviation at each time step. This situation corresponds to a longer period of oscillation, as required to observe a sufficient population of $\ket{\psi_4^+}$. To achieve this precision, we keep a rescaled time step $\Delta t  \sqrt{m^2 + J^2} = 0.1$, followed by a subsampling of the optimized circuit sequence, one for every $7$ steps, such that we are able to perform $125$ ``coarse-grained'' Trotter steps instead of $875$ for computational time reasons. We observe that the directly coupled $\ket{\psi_1^+}$ and $\ket{\psi_2^+}$ undergo a much faster oscillation, even if DQPTs are related with the slower one involving the resonant $\ket{\psi_4^+}$, thus yielding the dynamical free energy behavior of panel (b) in Fig.~\ref{fig:rabi_dqpt}. 

The statistics of simulations for this resonant condition of oscillating states are given in the lower panel of Fig.~\ref{fig:boxplot}. Fluctuations become of the same magnitude of resonant states shown in the other panels, while the detuned $\ket{\psi_3^+}$ still shows median values comparable with those characterizing the statistics of the middle panel. In this richer oscillating dynamics, each coupled state experiences a wider statistics for fluctuations. Any noise model yields the same statistics for each $\sigma^{\psi_i^+}_t$ of aforementioned coupled states, as opposed to the higher distinguishability characterizing the high detuning condition of the top panel.

\section{Conclusion and outlook} \label{conclusion}

The simulation of real-time dynamics studied in this paper is able to test Rydberg atom platform capabilities in the observation of DQPTs. By combining circuit depth compression in \texttt{qiskit} with pulse sequence engineering in \texttt{pulser} based on duration minimization, we implemented long time evolution with an arbitrary number of Trotter steps. In this way, we targeted a parameter regime in which the initial Dirac vacuum undergoes a resonant Rabi oscillation with mesonic states expected to represent the ground state for a negative mass parameter. We proved that, interestingly, multiple DQPTs are expected even with a small number of sites and that they can actually be observed on a real atomic platform, as shown by our simulations, which take into account a full combined noise model available in \texttt{pulser}. The scaling of the system will be considered in a future work, in order to also detect effective topological DQPTs.

\begin{acknowledgments}

We acknowledge support from INFN through the project ``QUANTUM'' and from PNRR MUR project CN00000013-``Italian National Centre on HPC, Big Data and Quantum Computing''.
FD, EE, PF, and DP acknowledge support from QCSC-Padova through the project ``Quantum Simulations 
for state preparation and dynamics of $\mathbb{Z} _3$-gauge Lattice Hamiltonians on Rydberg''.
EE, SP and FVP acknowledge support from PNRR MUR project PE0000023-NQSTI.
PF, SP, DP and FVP acknowledge support from the Italian funding within the ``Budget MUR - Dipartimenti di Eccellenza 2023--2027''  - Quantum Sensing and Modelling for One-Health (QuaSiModO).

\end{acknowledgments}

\bibliography{sn-bibliography}

\end{document}